\documentstyle[12pt,epsf]{article}
\newcommand{\beq}{\begin{equation}}
\newcommand{\eeq}{\end{equation}}
\newcommand{\beqa}{\begin{eqnarray}}
\newcommand{\eeqa}{\end{eqnarray}}

\newcommand{\barr}[1]{\not\mathrel #1}

\begin{document}

\noindent Accepted for publication in Phys. Lett. {\bf B}\\
 
{\bf rerevised version} \hfill KFA-IKP(TH)-1997-14

\hfill nucl-th/9707029

\vspace{1in}

\begin{center}

{{\Large\bf The reactions $pp\to pp\pi^0$ and $pp\to d\pi^+$ 
at threshold:\\[0.3em]
The role of the isoscalar $\pi N$ scattering amplitude}}

\end{center}

\vspace{.3in}

\begin{center}
{\large
C. Hanhart\footnote{email: kph168@ikp187.ikp.kfa-juelich.de},
J. Haidenbauer\footnote{email: J.Haidenbauer@fz-juelich.de},
M. Hoffmann\footnote{Present address: Commerzbank AG, Frankfurt, 
Germany},\\[0.2cm] 
Ulf-G. Mei{\ss}ner\footnote{email: Ulf-G.Meissner@fz-juelich.de},
J. Speth\footnote{email: J.Speth@fz-juelich.de}}

\bigskip

{\it Forschungszentrum J\"ulich, Institut f\"ur Kernphysik 
(Theorie)\\ D-52425 J\"ulich, Germany}

\bigskip

\end{center}

\vspace{.4in}

\thispagestyle{empty} 

\begin{abstract}\noindent
We examine the role of the elementary isoscalar pion--nucleon
scattering amplitude in the description of the processes $pp \to
pp\pi^0$ and $pp\to d\pi^+$ at threshold. 
We argue that the presently used tree level dimension
two approximation used in chiral perturbation theory is insufficient
as input by direct comparison with the $\pi$N scattering data.
We also show that a successful semi--phenomenological boson--exchange
model does better in the description of these data. 
The influence of the violation of crossing symmetry in the 
meson--exchange model has to be studied in more detail. We stress that
further investigations of the process $pp\to d\pi^+$ can pave the way to 
a deeper understanding of the pion dominated part of the transition
operator.

\end{abstract}

\vfill

\pagebreak

\noindent {\bf 1.} The high precision data for the processes $pp \to  
pp\pi^0$ and $pp\to d\pi^+$ in the threshold
region~\cite{meyer}\cite{who} have spurred a flurry of theoretical
investigations. The first data on neutral pion production were a big 
surprise because the experimental cross sections turned out to be a factor of 
five larger than the theoretical predictions. These included the direct
(Born) graph (fig.1a) and the on--shell rescattering (fig.1b), with the
pion--nucleon ($\pi N$) 
T--matrix replaced by the scattering length. Subsequently, it was argued
that heavy--meson exchanges (fig.1c) might be able to remove this 
discrepancy~\cite{LR}.  On the other hand it was shown~\cite{oset}\cite{unsers}
that the off--shell behaviour of the full $\pi N$ T--matrix also
enhances the cross section considerably. 
In \cite{unsers}, the pion--nucleon T--matrix was calculated within a 
semi--phenomenological meson--exchange model, based on an effective
Lagrangian fullfilling certain requirements from chiral symmetry (as
discussed below). Recently this 
result has been questioned based on calculations within the
framework of tree--level chiral perturbation theory (CHPT)
including dimension two operators~\cite{pmmmk,bira1,bira2,lee}.
In these papers, CHPT has been used to constrain the
long--range pion--exchange contributions.
 In fact, the chiral
perturbation theory approach, which abides to all symmetry
requirements including crossing\footnote{Note, however, that at
tree level unitarity is violated since all amplitudes are real.}, 
is believed to lead to a deeper
understanding of the success of the meson--exchange picture, as first
stressed by Weinberg~\cite{wein}. {\it It is therefore striking that the
calculations  for $pp \to pp\pi^0$
performed so far lead to a marked difference in the
role of the so--called rescattering contribution}, which interferes
constructively with the direct production in the J\"ulich model and 
destructively in the chiral framework, respectively. Note that while there is
still debate about the actual numerical treatment (co-ordinate versus
momentum space) and the ensuing size of the rescattering contribution
in the chiral perturbation theory approaches~\cite{lee}, the sign
difference to the meson--exchange model can be considered a genuine
feature. It is exactly this point which we wish to address in this
letter. We argue that the treatment underlying the isoscalar
pion--nucleon scattering amplitude and the related transition operator
for the process $NN \to NN\pi$ in the chiral framework is not yet 
sufficiently accurate and thus the resulting rescattering contribution
should be considered an artifact of this approximation. Clearly, this 
does not mean that chiral perturbation theory is invalid but rather
that higher order  (one loop) effects need to be accounted for. This
statement can be made very transparent if one considers in detail
the isoscalar $\pi N$ scattering amplitude. We stress, however,
that in the meson--exchange model crossing symmetry is violated and
we can not exclude the possibility that this is in fact the reason
for the constructive interference. To gain a deeper understanding
of the underlying direct and rescattering contributions, we propose
to study in more detail the process $pp\to d\pi^+$ since heavy meson
exchanges play no role in that case and one thus has a better chance
to extract the pionic (chiral)  physics which underlies the parameterization of
the $NN\to NN\pi$ transition operator. We furthermore mention that for
our investigation mainly focusing on the strength of the S--wave
rescattering, the explicit inclusion of the $\Delta (1232)$ is not
needed. Clearly, for a precise comparison with the data, this
approximation can not be maintained. Some of the effects of the
$\Delta$ are encoded in the numerical values of the low--energy constants
$c_i$. Further dynamical effects related to the $\Delta$
 only appear at higher orders togther
with many other terms. A systematic inclusion of such
effects goes beyond the scope of this letter. We stress that our
intention is to critically assess the accuracy of the presently
available CHPT calculations for pion production in $pp$ collisions.

\medskip

\begin{figure}[bht]
   \vspace{0.9cm}
   \epsfysize=5cm
   \centerline{\epsffile{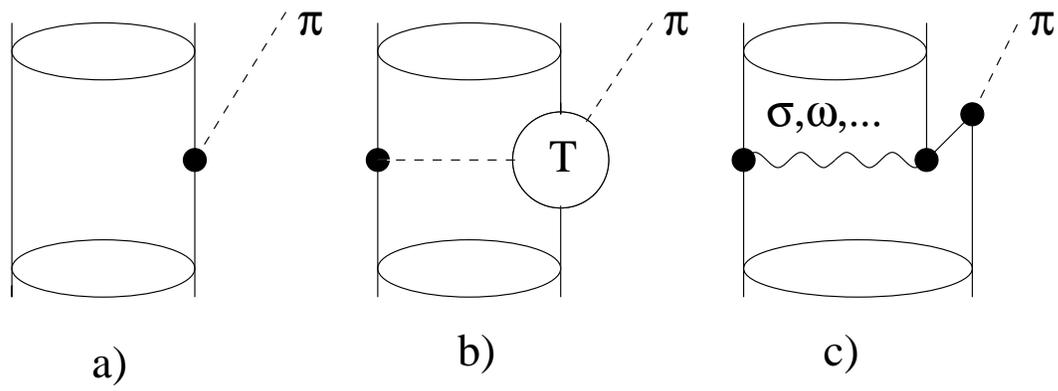}}
   \vspace{2cm} 
   \centerline{\parbox{15cm}{\caption{\label{fig3}
Different contributions considered for the meson production.
a), b) and c) are referred to as the direct, rescattering and 
heavy meson--exchange terms, in order. NN interactions are depicted
by the blobs.  }}}
\end{figure}

\noindent The important feature of this reaction is the large momentum mismatch
between the initial and the final nucleon--nucleon state: the
initial relative momentum is given by $p_{in}^2 =  mM_\pi + {M_\pi^2}/{4}$,
whereas close to threshold the final one is compatible with zero. This
huge difference leads to momentum transfers in the order of $p_{in}$, at
least if one  restricts oneself to S--wave production.
Let us concentrate first on the so--called rescattering contribution depicted
in fig.~\ref{fig3}b. To get a feeling for the relevant momentum transfer,
consider the reaction at threshold and neglect the $pp$ final--state interactions.
One finds for the four--momentum  of the
exchanged pion, $k^2 = -mM_{\pi} \simeq -7M_\pi^2 \simeq -0.1\,$GeV$^2$, 
which is deeply space--like. We remark that in single pion electroproduction
off protons, one--loop effects for such photon virtualities are fairly
large~\cite{bklm} substantiating our previous statement that one 
has to go beyond tree level in CHPT. 
\medskip

\noindent {\bf 2.} The study of pion--nucleon scattering in chiral
perturbation theory  has already a long history. In the relativistic
approach, Gasser et al.~\cite{gss} constructed the full off-shell $\pi$N
amplitude to one loop in the chiral expansion. 
Their work focused mainly on the determination
of the pion--nucleon $\sigma$--term and the related scattering lengths. A
systematic comparison with the low energy $\pi$N scattering data was,
however, never presented. 
In the framework of heavy baryon CHPT, the chiral corrections to the
S--wave scattering lengths were worked out by Bernard et al.~\cite{bkma}. 
In particular, it was shown that the isoscalar scattering 
length $a^+$ can be expressed as (to one loop)
\beq 
4\pi \biggl( 1 + \frac{M_\pi}{m} \biggr) \, a^+ =
\frac{2M_\pi^2}{F_\pi^2} \biggl( c_2 + c_3 -2c_1 - \frac{g_A^2}{8m}   
 \biggr) + \frac{3 g_A^2 M_\pi^3}{64 \pi F_\pi^4} + {\cal O}(M_\pi^4)
\,\,\, ,
\eeq
with $M_\pi$ ($m$) the charged pion (nucleon) mass, $F_\pi = 92.4\,$MeV
the pion decay constant and $g_A = 1.26$ the nucleon axial--vector 
coupling constant.  The $c_i$ are low--energy constants (LECs) not fixed
by chiral symmetry that appear in the dimension two chiral Lagrangian,
\beq\label{LpiN2}
{\cal L}_{\pi N}^{(2)} = \bar N \, \biggl( c_1 \, {\rm Tr}(\chi_+)
+ \bigl(c_2 - {g_A^2\over 8m}\bigr) (v \cdot u)^2 + c_3 \, u\cdot u 
+ \ldots \biggr) \, N  \,\,\, ,
\eeq
where the ellipsis denotes further terms not needed here, $u_\mu$ is
the standard axial--vector, which transforms homogeneously under
non--linearly realized chiral symmetry, $v_\mu$ the nucleons' 
four--velocity ($v^2 = 1$), $\chi_+ \sim M_\pi^2$ parameterizes the explicit
chiral symmetry breaking due to the light quark masses (and thus is
directly related to the $\sigma$--term) and $N$ is the velocity--projected
nucleon isodoublet, $N = (1 + \barr v ) \Psi /2$  
(for a detailed review, see~\cite{bkmrev}). The $c_i$ are finite since loop
corrections start to  contribute only at the next order in the chiral
expansion. The
precise values of these LECs will be discussed below. It is important
to note that there are large cancelations between the one loop
contribution of order $M_\pi^3$ and the kinematically $1/m$
corrections at order $M_\pi^2$. This makes the actual value of $a^+$
{\it very} sensitive to the  values of the LECs $c_i$. 
The scattering amplitude in the tree approximation to second order
in small momenta is readily derived from Eq.(\ref{LpiN2}) since to leading
order in pion fields, $u_\mu = -i \partial_\mu \phi / F_ \pi + 
{\cal O}(\phi^2)$. This is the accuracy to which this amplitude has been
used in chiral perturbation theory approaches to $pp \to pp\pi^0$ in the
threshold region. In that case, the
produced pion is almost at rest, i.e. has a very small three--momentum 
(denoted $q$), whereas the exchanged pion typically carries a momentum $k$ of
a few hundred MeV. The explicit form of the half--off shell scattering
amplitude $T^+ (q,k)$ is e.g. given in \cite{pmmmk}. Clearly, the sensitivity
to the precise values of the LECs observed for the scattering length carries
over to the scattering amplitude. Note that this isoscalar scattering
amplitude gives rise to the so--called rescattering contribution.

To proceed, we briefly summarize what is known about the LECs $c_i$. First,
it is important to notice that there is an ambiguity in pinning down their
values. These depend on the order one is working. From a tree level fit to 
pion--nucleon scattering (sub)threshold parameters including the dimension
two operators, one obtains~\cite{bkmppn} 
\beq\label{citree}
c_1 =  -0.64 \pm 0.14 \, \, , \quad
c_2  =  1.78 \pm 0.10 \, \, , \quad 
c_3  = -3.90 \pm 0.09 \, \, ,
\eeq
with all numbers given in GeV$^{-1}$. Note that the sum $c^+ =  
c_2 + c_3 -2c_1$ which enters the isoscalar amplitude, is $c^+ = -0.84$.
In this determination, the empirical value for $a^+$ was not used simply
because it is badly determined and also, it is not possible to describe
all (sub)threshold data that are sensitive to the $c_i$ by one simultaneous
fit at order $M_\pi^2$
(as discussed in detail in \cite{bkmci}). We therefore have also performed
calculations with $c^+= 0.459$ and $0.005$~GeV$^{-1}$ corresponding to
the the conservative band of $a^+ = \pm 10 \cdot 10^{-3}/M_{\pi}$,
respectively. Since there
are no compelling arguments which of the $c_i$ should be readjusted, we
have changed either one of them so as to get these values for $a^+$.

To one loop order, one can determine
these LECs from a set of seven observables which are given by tree graphs
including the $c_i$ and finite loop corrections but have no contribution
from the 24 dimension three LECs~\cite{bkmci}. One finds 
\beq\label{ciloop}
c_1 =  -0.93 \pm 0.10 \, \, , \quad
c_2  =  3.34 \pm 0.20 \, \, , \quad 
c_3  = -5.29 \pm 0.25 \, \, .
\eeq
The differences between the numbers given in Eq.(\ref{citree}) and
Eq.(\ref{ciloop}) are sizeable showing that the one--loop corrections
are not small and thus have to be taken into account. We remark that
in this case $c^+ = -0.09$ and $a^+ = -4.7 \cdot 10^{-3}/M_\pi$, well
within the empirical band.
Note that $c^+$ is an order of magnitude smaller than any of the 
LECs individually and is well within the uncertainty of the LECs. This
observation is at the origin of the statement that to understand the 
empirical value of  $a^+$, one has to know these LECs very precisely.
Mojzis~\cite{mm} has recently performed a complete one--loop calculation 
of the
$\pi$N scattering amplitude, including the dimension three counterterms.
A fit to the known S-, P-, D- and F-wave threshold parameters as well
as to the pion--nucleon $\sigma$--term and the Goldberger--Treiman 
discrepancy allows to pin down all LECs.
Remarkably, the dimension two LECs of relevance here come out completely
consistent with the values found in~\cite{bkmci}.    
To assess the sensitivity to the LECs, we will use both sets even so
we will only work to second order in the chiral expansion. 
This was also done in \cite{pmmmk}\cite{bira2}.
At this point, one might already suspect that this approximation is
not sufficient.

\begin{figure}[h]
   \vspace{0.9cm}
   \epsfysize=15cm
   \centerline{\epsffile{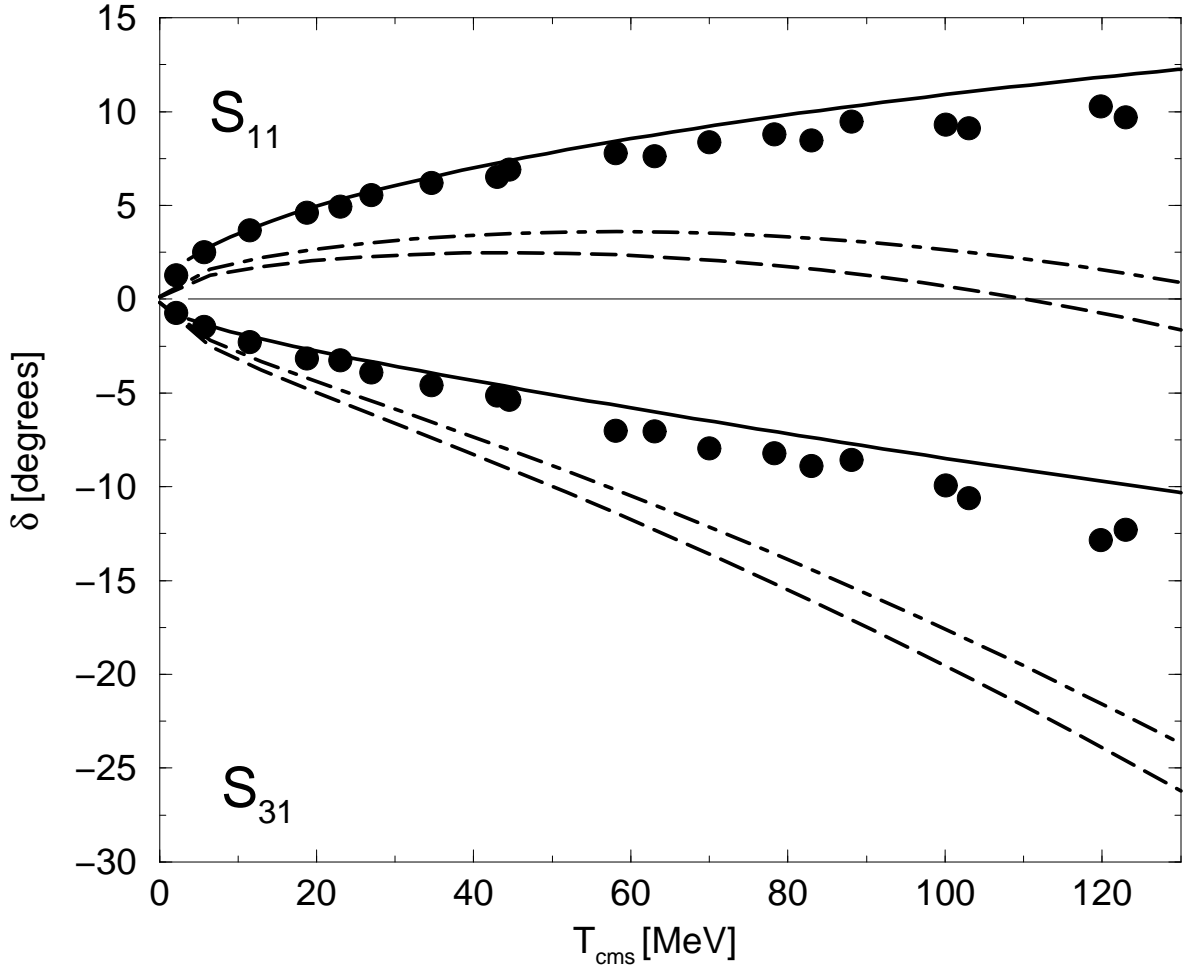}}
   \vspace{2cm} 
   \centerline{\parbox{15cm}{\caption{\label{fig2}
Comparison of the $\pi N$  phase shifts $S_{11}$ and $S_{31}$ 
for the meson--exchange model (solid lines) and the tree level 
CHPT calculation based on  parameter sets 1, Eq.(3) (dashed line) 
and 2, Eq.(4) (dot--dashed lines) with the data.
  }}}
\end{figure}

In contrast, the isovector amplitude is dominated by the Weinberg
term, i.e. the $\pi\pi \bar N N$ vertex which stems from the chiral covariant
derivative and has it strength given entirely in terms of $1/4F_\pi^2$.
The chiral corrections have also been calculated~\cite{bkma}\cite{bkmq4},
but since they are small, they are not important for the following arguments.
By forming appropriate linear combinations of the isoscalar
and isovector amplitudes, one can construct the phase shifts $S_{11}$ 
and $S_{31}$. They consist of a Born and a non--Born piece and take
the form (expanding the partial waves, not the invariant amplitudes,
to second order in small momenta)
\beqa
S_{11} &=& \sqrt{\omega^2 - M_\pi^2} \, \biggl( 1 - {\omega \over m} \biggr)
{1 \over 4\pi F_\pi^2} \biggl\{ -\omega + {1\over 2m} (M_\pi^2 -
\omega^2) + 2 M_\pi^2 c_1 -2\omega^2(c_2+c_3) \nonumber \\
&& \qquad\qquad\qquad\qquad\qquad\qquad
 +{g_A^2 \over 4} \biggl( {3 \omega^2\over 2m} - {1\over 3m}
\biggl[ 4M_\pi^2 + {M_\pi^4\over \omega^2} - {7\omega^2\over 2}\biggr] 
\biggr) \biggr\}  
 \,\,\, ,\nonumber \\
S_{31} &=& \sqrt{\omega^2 - M_\pi^2} \,  \biggl( 1 - {\omega \over m} \biggr)
{1 \over 4\pi F_\pi^2} \biggl\{ {\omega \over 2} - {1\over 4m} (M_\pi^2 -
\omega^2) + 2 M_\pi^2 c_1 -2\omega^2(c_2+c_3) \nonumber \\
&& \qquad\qquad\qquad\qquad\qquad\qquad
 +{g_A^2 \over 6m} 
\biggl[ 4M_\pi^2 + {M_\pi^4\over \omega^2} - {7\omega^2\over 2}\biggr] 
\biggr\}   \,\,\, ,
\eeqa
with $\omega$ the pion cms energy. The phase shifts 
are shown in fig.~\ref{fig2} for the $c_i$ as given in 
Eqs.(\ref{citree},\ref{ciloop}) in comparison to the data. In both
cases, the prediction deviate already at low energies from the data.
Due to the approximations involved, in both cases the partial waves increase
quadratically at higher energies, i.e. one has to include loop effects if
one wishes to find a good representation of the $\pi$N phase shifts for
energies up to 100~MeV or higher. We have also calculated these phases 
at tree level (dimension two) with $c^+$ adjusted so as to give the
empirical range of $a^+$ as discussed above. Enhancing e.g. $c_2$ by
a factor of 1.7 in the tree level fit leads to a good
description of the phase shifts up to $T_{\rm cms} \simeq
125$~MeV. Such an enhancement is, however, incompatible with some
subthreshold parameters, for detailed study see~\cite{bkmci}. We
therefore do not consider this a justified way of describing elastic
pion--nucleon scattering for the energies under consideration.
Furthermore,  at higher energies the unphysical quadratic growth
shows up again. We remark that the behavior of the phase shifts shown has
direct consequences for the transition operator relevant for $NN \to
NN\pi$ as discussed below.\medskip

\noindent {\bf 3.} There is a different way of constructing the $NN \to NN\pi$
transition operator that is by means of a meson--exchange model as
e.g. the one constructed by Sch\"utz et al.~\cite{Sch}. We briefly
summarize its salient features without going into any kind of 
detail (see also the discussion in ref.\cite{unsers}). 
The pseudo--potential contains in addition to the nucleon-- and $\Delta$--pole 
and exchange diagrams $\rho$ and $\sigma$ 
t--channel--exchanges. However, the latter two do not appear
as sharp particles but are constructed from the correlated two--pion
exchange by use of a dispersion integral.
The potential is then iterated in a relativistic Lippmann--Schwinger
equation within the framework of time--ordered perturbation theory
to ensure unitarity. To get finite results, the vertices are supplemented
by meson--nucleon form factors. This approach gives a unique
prescription to go off shell.
The parameters are then fitted to the $\pi N$ phase shifts up to 
lab energies of about 500~MeV under the constraint that the scattering
lengths come out to be in the empirical ranges.
In addition it turns out that the model produces a reasonable value
for the $\pi N$ $\sigma$--term.
As it was stated above the value of the isoscalar scattering length
is compatible with zero. In the meson--exchange model, one finds
$a^+ = -1.7 \cdot 10^{-3}/M_\pi$. This small value is a consequence
of a cancelation between the $\sigma$--exchange and the iterated 
$\rho$--exchange.\footnote{Note that on the potential level, the $\rho$ only 
contributes to the isovector channel.} 
However, a priori there is no reason for this cancellation to remain
valid when leaving the on--shell point. We actually observe a strong
momentum dependence in the half--off-shell isoscalar T--matrix at
threshold. The sign of this function is fully determined by the
fit to the on--shell data.
The drawback of this approach is the blatant violation of crossing
symmetry which is due to the iteration in the Lippmann--Schwinger
equation to generate the full T--matrix.  To be precise, the
iteration generates contributions with powers in $\nu$ which should
be zero if crossing were to be respected. Some of those coefficients
turn out to have non--negligible strength. However, one can 
calculate the subthreshold parameters, i.e. the expansion
of the invariant amplitudes around the point $\nu = t = 0$ in the
$\nu-t$ plane, which are allowed by crossing. For doing that, we used
for convenience model~1 of ref.\cite{Sch}.
These are tabulated in table~1 (for definitions, see ref.~\cite{bible}).

\renewcommand{\arraystretch}{1.4}
\begin{table}[bht] 
\begin{center}

\begin{tabular}{|c||c|c||c|c||c|c||}
    \hline
     Amp. & $x_{00}$ & Exp. &  $x_{01}$ & Exp. &   $x_{02}$ & Exp. \\ 
    \hline
    $A^+$     & $-1.54$  & $-1.46\pm 0.10$ & $1.20$ & $1.14 \pm 0.02$ &
                $0.018$  & $0.036 \pm 0.003$   \\    
    $A^-/\nu$ & $-11.46$ & $-8.83\pm 0.10$ & $-0.28$ & $-0.374 \pm 0.002$ &
                $-0.028$   & $-0.015 \pm 0.002$   \\    
    $B^+/\nu $& $-2.6$  & $-3.54\pm 0.06$ & $0.13$ & $0.18 \pm 0.01$ &
                $-0.004$   & $-0.01$   \\    
    $B^-$     & $12.64$  & $10.36\pm 0.10$ & $0.20$ & $0.24 \pm 0.01$ &
                $-0.02$  & $0.025 \pm 0.002$   \\    
          \hline
  \end{tabular}
\end{center}
\caption{
Subthreshold parameters calculated in the meson--exchange model after
neglecting the crossing--violating terms (which affects in
particular $A^-/\nu$ and $B^+/\nu$). The data are from
ref.~{\protect{\cite{bible}}} and the units are appropriate powers of
the inverse pion mass.}
\end{table}

\medskip

\noindent {\bf 4.} We now turn to the reaction $pp \to pp\pi^0$. 
The pertinent transition
 matrix element is calculated in the distorted wave Born approximation,
 allowing us to properly include
the final state nucleon--nucleon interaction. This was shown to be
crucial to get the correct energy--dependence of 
the reaction under consideration~\cite{MuS}.

\begin{figure}[h]
   \vspace{0.9cm}
   \epsfysize=15cm
   \centerline{\epsffile{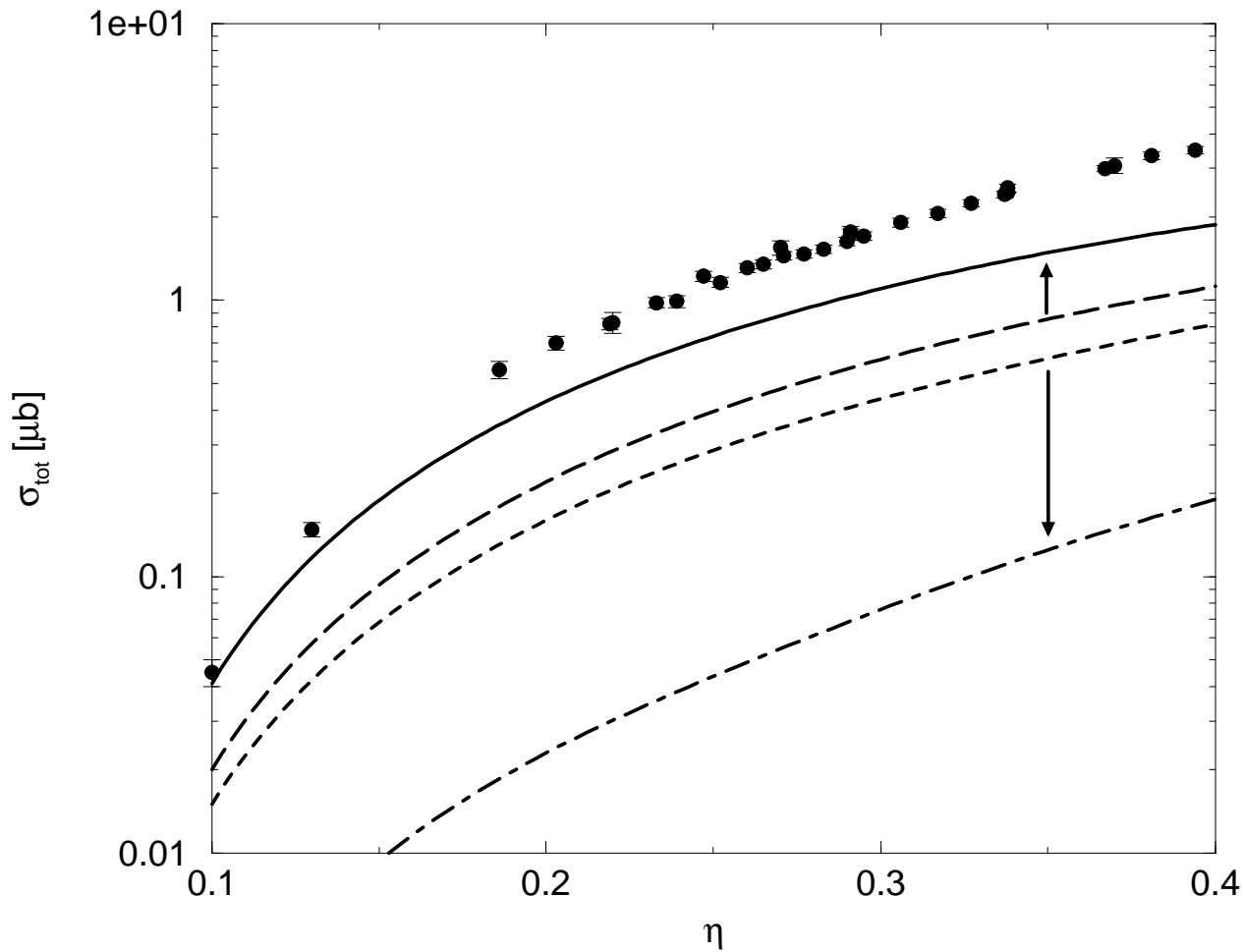}}
   \vspace{1.5cm} 
   \centerline{\parbox{15cm}{\caption{\label{fig4}
Results for S--wave $pp \to pp \pi^0$. 
The long--dashed line is the result of the 
rescattering contribution from the meson exchange--model. Adding the
direct contribution to this gives the upper solid line (both interfere
constructively as indicated by the arrow).
The short--dashed line is the result of the 
rescattering contribution from tree level CHPT (parameters from
   Eq.(3)).  Adding the
same amplitude for the direct contribution as before leads to
the lower solid line (the arrow indicates the destructive interference).
Note that heavy meson exchanges are not included. $\eta$ is the maximal
pion momentum in units of the pion mass. 
}}}
\end{figure}

As it was stated above, it is only the isoscalar T--matrix that enters in this
reaction. It turns out, however, that the results of the two approaches 
described before are very different. The chiral tree calculation leads
to a destructive interference with respect to the direct contribution 
(cf. fig.~\ref{fig3}a) (see e.g. \cite{pmmmk}), whereas in
the meson--exchange model one finds a constructive interference~\cite{unsers}
(the main difference of the curves presented here to the ones in \cite{unsers}
is that now the backward--in--time rescattering 
is considered in addition; details will
be described elsewhere \cite{prep}).
The results for the different approaches are shown in fig.~\ref{fig4}
(using the LECs given in Eq.(\ref{citree}) for the chiral calculation).
In the numerical treatment of the chiral tree level approach we deviate
from the strict CHPT treatment of the production operator to the extent that
we apply a form factor on the pion--nucleon vertex. This was done not only 
to achieve better convergence of the integral but also to prevent the
production operator to show the abovementioned unphysical linear growth
in the  momentum.
Although this treatment changes the quantitative results it does not effect
their {\it  qualitative} behavior.
We have checked that while the actual strength of the rescattering
contribution in CHPT is very sensitive to the values of the LECs
$c_i$, its sign is uniquely fixed to this order (dimension two).
In both results there is a remaining discrepancy to the data.
A possible mechanism to fill this gap are the so called heavy meson
exchanges~\cite{LR}. However, their sign is such that they give a 
constructive interference with the direct term (actually allowing
for a good description of the data \cite{prep}). Heavy meson exchange 
will therefore even worsen the discrepancy for the tree level chiral
calculation~\cite{lee}.

\begin{figure}[h]
   \vspace{0.9cm}
   \epsfysize=15cm
   \centerline{\epsffile{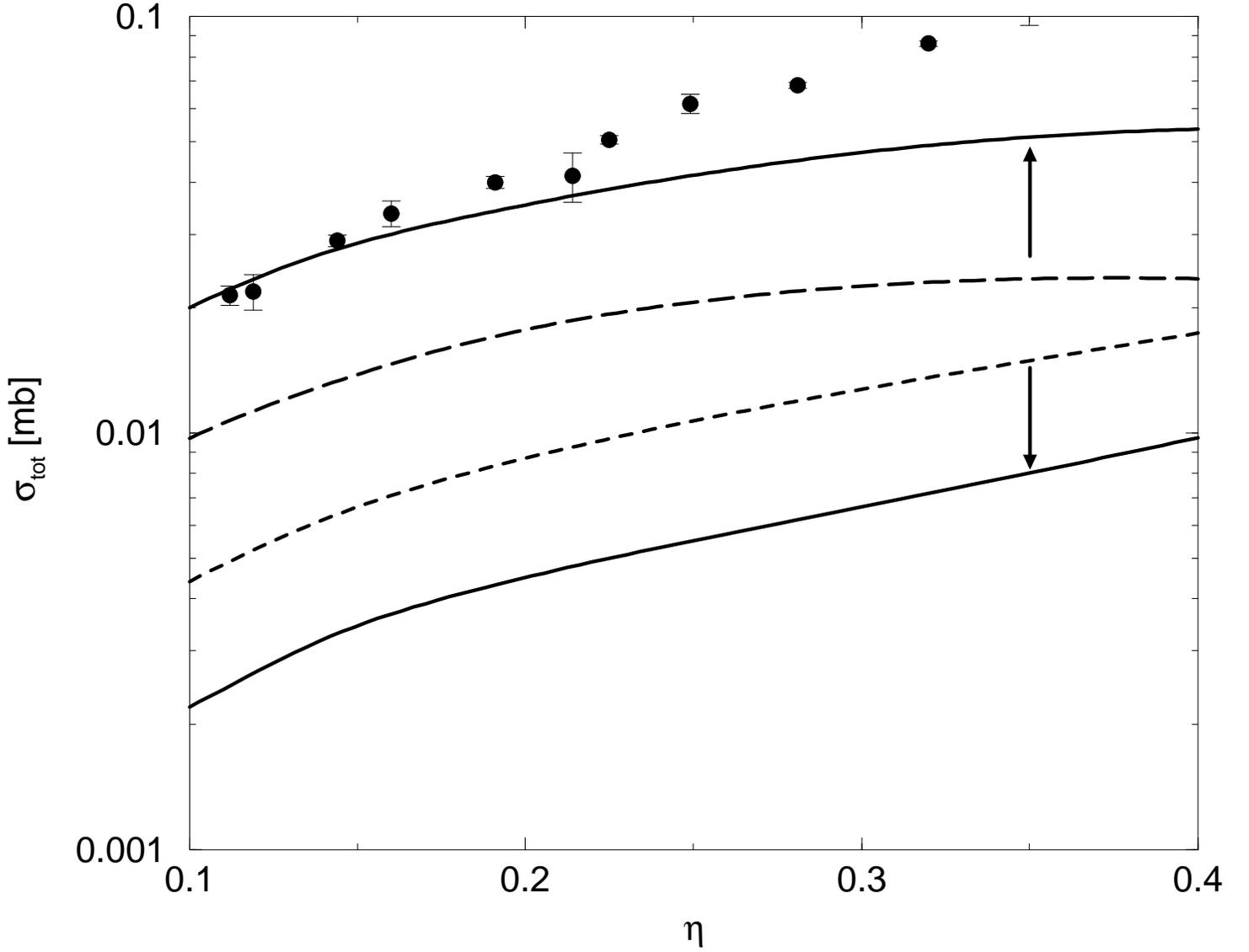}}
   \vspace{1.5cm} 
   \centerline{\parbox{15cm}{\caption{\label{fig5}
Influence of the isoscalar $\pi N$ amplitude on S--wave $pp \to d \pi^+$. 
The long--dashed line shows the isovector rescattering of 
the meson--exchange model.
Adding the isoscalar contribution leads to the upper solid line.
The short--dashed line is the isovector rescattering contribution
of the tree level CHPT. Adding the CHPT--isoscalar amplitude
to this leads to the lower solid curve, which substantially deviates
   from the data.
Note that the effects of heavy meson exchanges and the direct term are
small and not included here. In this case, $\eta = |\vec{q} \,|/ M_\pi$.
 }}}
\end{figure}

We also applied for the first time the CHPT approach to the reaction $pp\to d \pi^+$. 
Here neither
the direct contribution nor the heavy meson exchanges contribute considerably
\cite{nis}. In this reaction a charge exchange is allowed and thus the much
larger isovector channel of the $\pi N$ system can contribute. To make the
comparison of the different approaches in this reaction, we show in 
fig.~\ref{fig5} the effect of including the isoscalar rescattering with respect
to the isovector contribution.
As before, the interference pattern of the two
 approaches is totally different. While  
in case of the meson--exchange picture the isoscalar rescattering
is  bringing the theory in agreement with
the data, using tree level chiral perturbation theory worsens the description.
We conclude from this comparison that more detailed studies of this
reaction might eventually lead to a deeper understanding of the
relative size of the isovector and isoscalar components entering the
$NN \to NN\pi$ transition operator.\medskip

\noindent {\bf 5.} To summarize, we have presented a detailed study
about the relative sign of the rescattering and direct terms for 
single pion production in $pp$ collisions in the framework of
meson--exchange and tree level
dimension two chiral perturbation theory calculations. While the latter
is a more fundamental approach, the underlying (small) isoscalar
scattering amplitude and the related $NN \to NN\pi$ transition
operator are not sufficiently accurate to the order they have been
treated so far for the momenta involved. We can make this statement
even  more precise. {\it Only at one loop order, terms proportional to
the squared momentum of the exchanged pion appear in the $\pi N$
amplitude. We expect exactly these terms to control the $\pi N$
T--matrix at large space--like momenta. Neglecting these terms
is most probably at the heart of the sign discrepancy discussed here.}
We stress again that
this does not mean that chiral perturbation theory can not be used,
but rather that one has to work out higher order (loop) effects before one
can draw decisive conclusions. In any case, a sufficiently improved
CHPT calculation will have to be supplemented by heavy meson exchanges
(or contact terms representing this short--range physics) to be able 
to describe the  $pp \to pp\pi^0$ data. The more successful
meson--exchange model, which describes the on--shell $\pi N$
scattering data and does much better for $pp \to pp\pi^0$, has the 
deficiency of violating crossing symmetry. One can therefore
not definitively conclude that this is the reason for the relative
sign difference to the chiral approach, although this appears to be
a fairly exotic possibility.  In that respect, more attention should be
focused on $pp \to d\pi^+$ since it is much less sensitive to the
heavy meson exchanges but still sensitive enough to the isoscalar  
scattering amplitude.

\medskip

\section*{Acknowledgements}

We thank Norbert Kaiser and Martin Mojzis for useful communications.


\newpage

\section*{Figures}

\vspace{3cm}

\end{document}